\documentclass[journal,10pt]{IEEEtran}
\usepackage{amsmath}
\usepackage{graphicx}
\usepackage{subfig}
\usepackage{color}
\usepackage{float}
\usepackage{cite}

\begin{document}

\title{Task-oriented Over-the-Air Computation for Edge-device Co-inference with Balanced Classification Accuracy}

\author{Xiang Jiao, Dingzhu Wen, Guangxu Zhu, Wei Jiang, Wu Luo, Yuanming Shi



\thanks{Xiang Jiao is with State Key Laboratory of Advanced Optical Communication Systems and Networks, School of Electronics, Peking University, Beijing, China, and Shenzhen Research Institute of Big Data, Shenzhen, China (e-mail: JiaoXiang@stu.pku.edu.cn). (Corresponding authors: Dingzhu Wen, Wei Jiang)}

\thanks{Dingzhu Wen and Yuanming Shi are with Network Intelligence Center, School of Information Science and Technology, ShanghaiTech University, Shanghai, China (e-mail: wendzh, shiym@shanghaitech.edu.cn).} 

\thanks{Guangxu Zhu is with Shenzhen Research Institute of Big Data, Shenzhen, China (e-mail: gxzhu@sribd.cn).} 

\thanks{Wei Jiang and Wu Luo are with the State Key Laboratory of Advanced Optical Communication Systems and Networks, School of Electronics, Peking University, Beijing, China (e-mail: jiangwei, luow@pku.edu.cn).} 

}

\maketitle

\begin{abstract}
Edge-device co-inference, which concerns the cooperation between edge devices and an edge server for completing inference tasks over wireless networks, has been a promising technique for enabling various kinds of intelligent services at the network edge, e.g., auto-driving. In this paradigm, the concerned design objective of the network shifts from the traditional communication throughput to the effective and efficient execution of the inference task underpinned by the network, measured by, e.g., the inference accuracy and latency. In this paper, a task-oriented over-the-air computation scheme is proposed for a multi-device artificial intelligence system. Particularly, a novel tractable inference accuracy metric is proposed for classification tasks, which is called minimum pair-wise discriminant gain. Unlike prior work measuring the average of all class pairs in feature space, it measures the minimum distance of all class pairs. By maximizing the minimum pair-wise discriminant gain instead of its average counterpart, any pair of classes can be better separated in the feature space, and thus leading to a balanced and improved inference accuracy for all classes. Besides, this paper jointly optimizes the minimum discriminant gain of all feature elements instead of separately maximizing that of each element in the existing designs. As a result, the transmit power can be adaptively allocated to the feature elements according to their different contributions to the inference accuracy, opening an extra degree of freedom to improve inference performance. Extensive experiments are conducted using a concrete use case of human motion recognition to verify the superiority of the proposed design over the benchmarking scheme. 
\end{abstract}


\IEEEpeerreviewmaketitle

\section{Introduction}

Edge artificial intelligence (AI) has been an emerging technique to provide various kinds of intelligent services to support many applications like auto-driving and remote health   \cite{letaief2021edge,shi2023task-orientedsurvey,Li2022DL,cao2021JSAC_optimized}. However, the realization of these intelligent services demands the deployment of well-trained AI models at the network edge and utilizes their inference capability for making human-like intelligent decisions. This gives rise to a new research topic, called edge inference \cite{zhang2021deep,Hua2021TGCN,yun2021cooperative,Shao2020CM,shi2020device,Lan2022TWC,Wen2022ISCC,wen2022task}. 
There are now three types of edge inference: on-device inference, which uses the device for the entire inference task, takes up a lot of computational resources on the device (see, e.g., \cite{zhang2021deep}). The second one is known as on-server inference which uses a server to perform the entire inference task completely (see, e.g., \cite{Hua2021TGCN}). It causes huge communication overhead and privacy leakage. Among others, the technique of edge-device co-inference, called edge split inference, has been the most popular architecture \cite{Lan2022TWC,yun2021cooperative,Shao2020CM,shi2020device,Wen2022ISCC,wen2022task}. It divides an AI model into two parts. One part with light size is deployed at the device for extracting a low-dimensional feature vector from the high-dimensional raw data. The remaining computation-intensive part is deployed at the server and utilizes the received feature vector from the device to complete the downstream inference task. By avoiding high-dimensional raw data transmission and offloading the intensive computation to the server, the edge-device co-inference can enjoy the advantages of enhanced communication and computation efficiency as well as preserving data privacy and thus is considered in the current work.

\begin{figure*}[!t]
	\centering	  
        \includegraphics[width=0.80\textwidth]{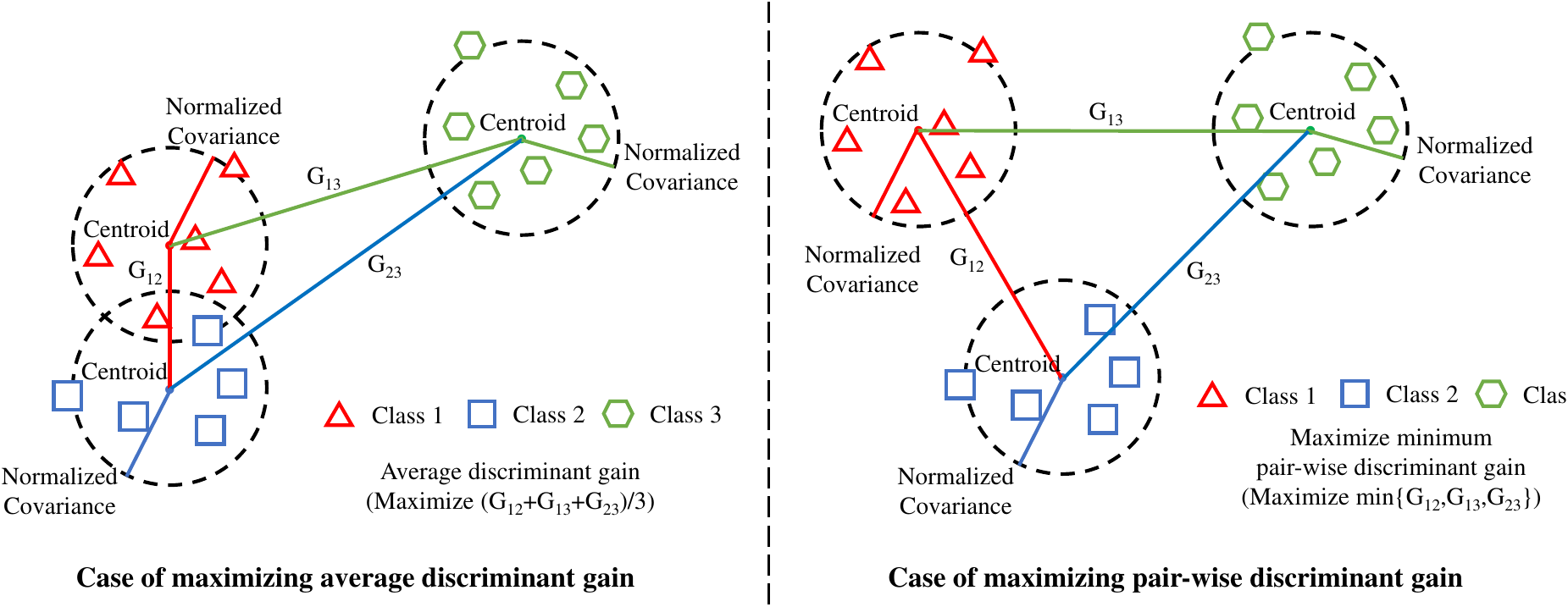}
	\caption{The difference between maximizing average discriminant gain and maximizing minimum discriminant gain.}
	\vspace{-0.7cm}
	\label{fig:DG}
\end{figure*}

As stated in \cite{Wen2022ISCC,wen2022task,zhang2024collaborative,wen2024integratedmagzine}, the design of edge-device co-inference calls for task-oriented communication techniques, since traditional techniques designed for throughput maximization or distortion minimization, fail to distinguish the data samples of different contributions on the inference task. To tackle this problem, a task-oriented scheme of integrated sensing, computation, and communication (ISCC) was proposed in \cite{Wen2022ISCC} for multi-device AI inference, where each device senses a target area from disjoint narrow views. Furthermore, for the case where different devices observe the same wide view of a target area with each providing a noisy observation, a task-oriented over-the-air computation (AirComp) was proposed in \cite{wen2022task} to efficiently aggregate the local features for suppressing the sensing and channel noise by directly adopting the inference accuracy as the design goal instead of using the traditional minimum mean square error (MMSE) (see, e.g., \cite{wen2019reduced, yang2023communication}). This task-oriented AirComp design was extended to the Cloud-RAN architectures \cite{zhang2024collaborative}. As the instantaneous inference accuracy is unknown at the design stage, when the input data of the AI model is not obtained, authors in \cite{Wen2022ISCC,wen2022task,zhang2024collaborative} adopted an approximate but tractable metric as a surrogate for classification tasks, called discriminant gain and originally proposed in \cite{Lan2022TWC} based on the well-known Kullback-Leibler divergence \cite{kullback1951information}. Specifically, as shown in Fig. \ref{fig:DG}, the discriminant gain of an arbitrary class pair (e.g., $G_{1,2},G_{1,3}$ or $G_{2,3}$) is defined as the distance between their centroids in the Euclidean feature space normalized by their covariance. With a larger discriminant gain, the two classes are better separated in the feature space, which implies a higher achievable inference accuracy.

However, the average discriminant gain metric adopted in  \cite{Wen2022ISCC,wen2022task, zhang2024collaborative} has a drawback of unbalanced inference accuracy of different classes, as shown in the left of Fig. \ref{fig:DG}. Specifically, with maximizing average discriminant gain, one particular class (e.g., class 3) may be far separated from the rest of the classes (e.g., class 1 and 2) which could be very close to each other in the feature space. As a result, only samples from one particular class (e.g., class 3) can be accurately predicted but those from other classes (e.g., class 1 and 2) cannot be distinguished. This leads to an unbalanced and low inference accuracy. To address this problem, this work proposes a novel metric based on pair-wise discriminant gain, whose goal is to maximize the minimum discriminant gain of all class pairs. Thereby, the worst separated class pair (e.g. classes 1 and 2) can receive better treatment, as shown in the right side of Fig. \ref{fig:DG}.

In this work, an edge-device co-inference system is considered with multiple single-antenna devices and a single-antenna edge server. All devices observe a common wide-view target source but each contributes a different noisy observation of the ground true source data. Then, each device extracts a low-dimensional local feature vector from the high-dimensional local sensory data. The AirComp technique is adopted to aggregate all local feature vectors to suppress the sensing and channel noise. Specifically, all devices transmit an element of the same feature dimension over the same resource block. At the server, they are aggregated automatically due to the waveform superposition property of the wireless channels and used to derive an estimate of the ground-true feature element. All elements are sequentially transmitted over multiple time slots. The criterion of minimum pair-wise discriminant gain maximization is adopted for improving inference accuracy. The detailed contributions are summarized as follows.

\begin{itemize}
    \item{\bf Inference Accuracy Enhancement via Maximizing Minimum Pair-wise Discriminant Gain}: This paper proposes a new metric, i.e., minimum pair-wise discriminant gain, to overcome the issue of unbalanced inference accuracy caused by the metric of average pair-wise discriminant gain. Under the new criterion, the class pair with the smallest distance in the feature space will receive better treatment, leading to a balanced and improved achievable inference accuracy.

    \item {\bf  Joint Optimization of All Feature Elements}: Instead of separately optimizing the average discriminant gain of each feature element in the existing design \cite{wen2022task}, this paper jointly optimizes the minimum pair-wise discriminant gain of all feature elements attained by multi-slot AirComp. Thereby, the transmit power can be adaptively allocated to different feature elements according to their different contributions to the inference accuracy, opening a new degree of freedom for performance improvement. To solve the resultant non-convex transmit power allocation problem, variable transformation is first applied to derive an equivalent difference-of-convex (d.c.) problem, which is then solved using the typical method of successive convex approximation (SCA). 
    
\end{itemize}

The structure of this paper is as follows. Section II introduces the system model and the problem formulation. Section III suggests minimum discriminant gain maximization. Section IV contains the performance evaluation, and Section V concludes the paper.

\section{System Model and Problem Formulation}

\subsection{System Model}

\begin{figure*}[!t]
	\centering	  
        \includegraphics[width=0.80\textwidth]{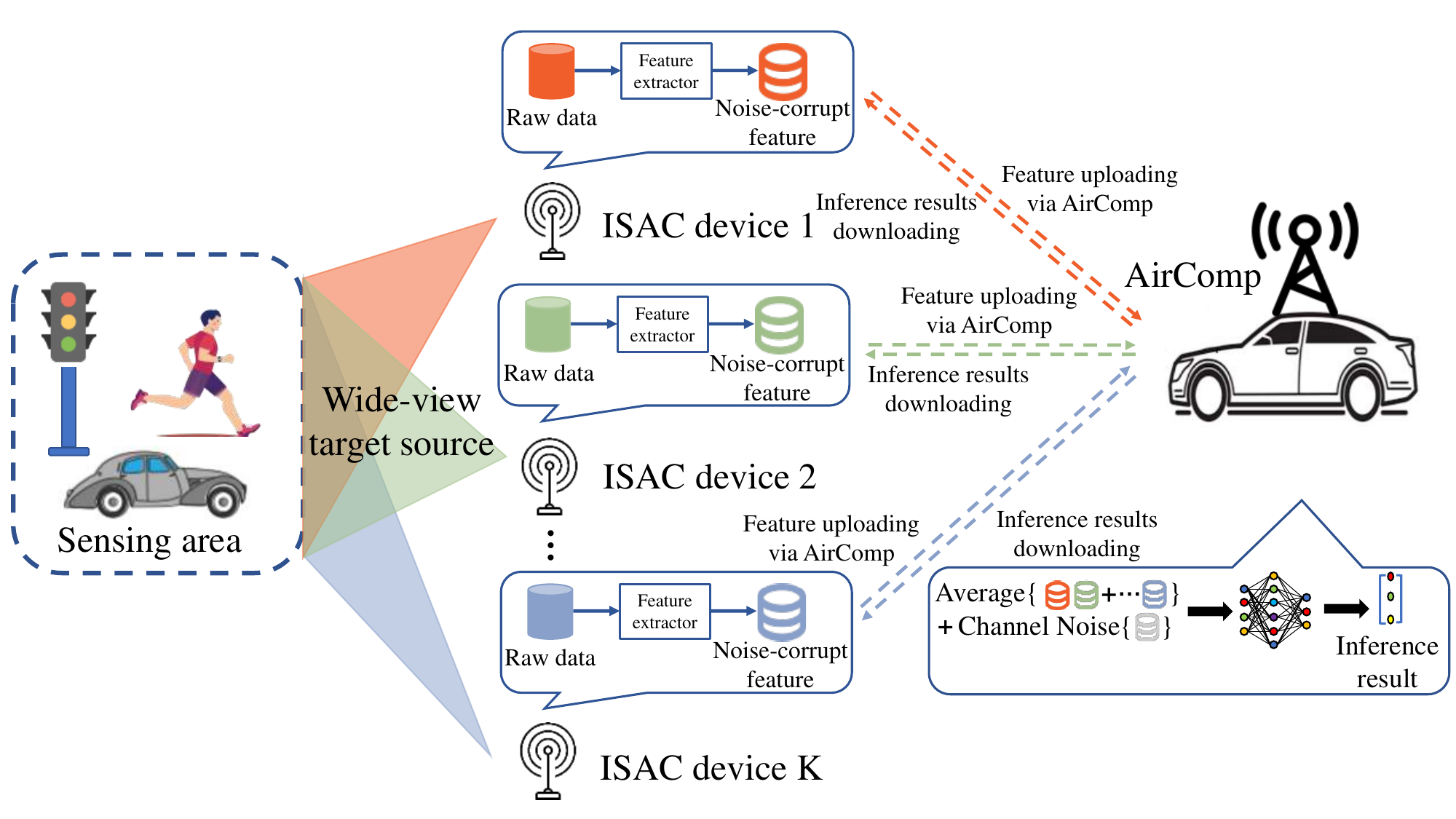}
	\caption{System model.}
	\vspace{-0.7cm}
	\label{fig:System_model}
\end{figure*}

\subsubsection{Network Model}
Consider an edge-device co-inference system with $K$ single-antenna devices and one edge server equipped with a single-antenna access point, as shown in Fig. \ref{fig:System_model}. The devices sense the same wide view of a target and obtain real-time noise-corrupted sensory data for the inference task. Then, a principal component analysis (PCA) based feature extractor is utilized to extract a local low-dimensional feature vector from the raw sensory data at each device. In practice, the number of extracted feature dimensions is task-dependent and can be pre-determined during the training stage. Next, for suppressing the sensing noise to achieve higher inference performance, all local feature vectors are aggregated at the server using AirComp to derive a denoised global one, which is further input into an AI sub-model for completing the downstream inference task \cite{wen2022task}. The number of extracted feature elements is denoted as $M$. As a reasonable example, the edge server can refer to a highly mobile vehicle on a road and the devices can refer to road site units \cite{Lan2022TWC}.

Time-division multiplexing (TDM) is used. At each time slot, an element of the same feature dimension is transmitted by each device and is aggregated at the server for denoising. The overall feature vector is sequentially transmitted over $M$ time slots. Since the transmission time of one element is far less than the coherence time \cite{Zhu2020TWC}, the channels are assumed to be static during the transmission of all $M$ elements. Particularly, the uplink channel gain of the $k$-th  device is denoted as $h_k$. Without loss of generality, the server works as the central coordinator which determines the transmit power for all devices and has the ability to acquire the channel state information of all involved links.

\subsubsection{Feature Distribution}
The extracted local feature vector is a noise-corrupted version of the ground-true one, as
\begin{equation}
    {\bf x}_k = {\bf x} + {\bf d}_k, \;\; 1\leq k \leq K,
\end{equation}
where $ {\bf x}= \{x_1,...,x_m,...,x_M\}$ is the ground-true feature vector, ${\bf d}_k=\{d_{k,1},...,d_{k,m},...d_{k,M}\}$ is the difference between the noise-corrupt feature and ground-true one which is assumed as the Gaussian feature sensing noise. As PCA is applied, different elements of ${\bf x}$ and ${\bf d}_k$ are independent. Consider a classification task with $L$ classes, 
the $m$-th element of ${\bf x}$ has a distribution of
\begin{equation}
    x_m \sim \frac{1}{L} \sum \limits_{\ell=1}^L \mathcal{N} \left({\mu}_{\ell,m}, \sigma_{m}^2 \right) ,
\end{equation}
where $\mathcal{N}\left({\mu}_{\ell,m}, \sigma_{m}^2\right)$ represents the Gaussian distribution corresponding to the $ \ell $-th class, ${\mu}_{\ell,m}$ is the mean of $m$-th dimension in the $ \ell $-th class, $ \sigma_{m}^2 $ is the variance of $m$-th dimension. In practice, the values of these parameters are estimated at the server in the training stage using the training data samples. The $m$-th element of ${\bf d}_k$ has a distribution of
\begin{equation}
    d_{k,m} \sim \mathcal{N} \left(0, \delta_{k,m}^2 \right), 
\end{equation}
where $\delta_{k,m}^2$ is the feature noise variance. It can be pre-estimated by first estimating the ground-true feature vector by averaging massive noisy ones and then analyzing the variance between each local feature vector and the estimated ground-true one. Thereby, the $m$-th element of ${\bf x}_k $, given by 
\begin{equation}
    x_{k,m} = x_m + d_{k,m},
\end{equation}
which has a distribution of
\begin{equation}\label{Eq:LocalFeatureDistribution}
    x_{k,m} \sim \frac{1}{L} \sum \limits_{\ell=1}^L \mathcal{N} \left({\mu}_{\ell,m},\sigma_{m}^2 +  \delta_{k,m}^2 \right).
\end{equation}

\subsubsection{AirComp}
To simultaneously access multiple devices to aggregate the local feature vectors for sensing and channel noise suppression, the technique of AirComp is adopted to reduce the communication overhead. For an arbitrary time slot $m$, all devices pre-code and transmit the $m$-th element of their local feature vectors, i.e., $x_{k,m}$ over the same resource block. Thereby, the received element at the server is given by
\begin{equation}
    \hat{x}_m = \sum \limits_{k=1}^K h_k b_{k,m} x_{k,m} + n,\;\; 1 \leq m \leq M,
\end{equation}
where $b_{k,m}$ is the precoding scalar at device $k$, $n$ is the channel noise with a distribution of $ n \sim \mathcal{N} \left(0,\delta_{0}^2 \right)$, and $\delta_0^2$ is the channel noise power. 

Then, by substituting the distribution of $x_{k,m}$ in \eqref{Eq:LocalFeatureDistribution} into $\hat{x}_m$, its distribution can be derived as
\begin{equation}\label{Eq:ReceiveFeatureDistribution}
    \hat{x}_m \sim \frac{1}{L} \sum \limits_{\ell=1}^L \mathcal{N} \left(\hat{\mu}_{\ell,m},\hat{\sigma}_{m}^2 \right),
\end{equation}
where $\hat{\sigma}_{m}^2 = \left(  \sum\limits_{k=1}^K h_k b_{k,m} \right)^2 \sigma_{m}^2 + \sum\limits_{k=1}^K h_k^2 b_{k,m}^2 \delta_{k,m}^2 + \delta_{0}^2$ and $\hat{\mu}_{\ell,m} = \sum\limits_{k=1}^K h_kb_{k,m}{\mu}_{\ell,m}$.

\subsubsection{Inference Accuracy Measured by Discriminant Gain}
In this work, inference accuracy is adopted as the design criterion instead of the traditional MSE in existing designs. Since the latter targets minimizing the distortion between the received feature vector and the ground-true one, it ignores that the same distortion level on different feature elements has different impacts on the inference accuracy \cite{Lan2022TWC,Wen2022ISCC,wen2022task}. However, the instantaneous inference accuracy is unknown in the design stage before the feature vector is inputted into the AI model. To address this issue, discriminant gain is adopted as the surrogate. Specifically, consider a classification task with $L$ classes, whose feature distribution is in \eqref{Eq:ReceiveFeatureDistribution}. For an arbitrary class pair, say classes $\ell$ and $\ell^{'}$, the discriminant gain between the two classes is 
\begin{equation}\label{Eq:PairDG}
    G_{\ell,\ell^{'}}({\bf x}) =  \sum\limits_{m=1}^M \left( \hat{\mu}_{\ell,m} - \hat{\mu}_{\ell^{'},m}\right)^2 / \hat{\sigma}_{m}^2.
\end{equation}
The pair-wise discriminant gain in \eqref{Eq:PairDG} represents the discernibility of the two classes in the feature space, as shown in Fig. \ref{fig:DG}. With larger discriminant gain, the two classes are better separated, leading to higher achievable inference accuracy.

\subsection{Problem Formulation}

Existing work adopts the metric of maximizing average pair-wise discriminant gain of all class pairs for enhancing the inference accuracy \cite{Lan2022TWC,Wen2022ISCC,wen2022task}. This, however, causes an unbalanced and low accuracy as mentioned before and as shown in the left side of Fig. \ref{fig:DG}. To address this issue, we propose a novel design criterion, which maximizes the minimum pair-wise discriminant gain over all class pairs. As a result, the class pair with the smallest distance can be well separated, resulting in enhanced achievable inference accuracy for all classes, as shown in the right side of Fig. \ref{fig:DG}. Besides, there are two kinds of constraints. One is the transmit power constraint for each device in each time slot. The other is the overall energy constraint of each device to transmit the whole feature vector. In summary, the problem is given by 
\begin{equation*}\text{(P1)}\;
\begin{aligned}
\max\limits_{\{b_k\}}  \min \limits_{ \left(\ell,\ell^{'}>\ell\right)} \;\; & G_{\ell,\ell^{'}}({\bf x}) =  \sum\limits_{m=1}^M {\left( \hat{\mu}_{\ell,m} - \hat{\mu}_{\ell^{'},m}\right)^2}/{\hat{\sigma}_{m}^2},
\\ \text{s.t.} \;\; & b_{k,m}^2 \leq P_k, \quad  \forall(k,m),
\\ & \sum \limits_{m=1}^M b_{k,m}^2 \leq \hat{P}_k,\; \forall k,
\end{aligned}
\end{equation*}
where $P_k$ is transmit power threshold in each time slot and $\hat{P}_k$ is total power constraint of all time slots. 

\section{Minimum Discriminant Gain Maximization}

(P1) is non-convex due to the complicated form of the objective function. To this end, the following variable transformation is applied to simplify (P1).   For all class pairs $ (\ell,\ell^{'}>\ell) $, we introduce 
\begin{equation}
    T_{m,\ell,\ell^{'}} \leq \left( \hat{\mu}_{\ell,m} - \hat{\mu}_{\ell^{'},m}\right)^2 / \hat{\sigma}_{m}^2,\quad \forall m.
\end{equation}
Then, denote $T = \sum\limits_{m=1}^M T_{m,\ell,\ell^{'}},\quad \forall (\ell,\ell^{'}>\ell)$.
Accordingly, it is easy to derive that
\begin{equation}
    T \leq \;\; \sum\limits_{m=1}^M \left( \hat{\mu}_{\ell,m} - \hat{\mu}_{\ell^{'},m}\right)^2 / \hat{\sigma}_{m}^2,\quad \forall (\ell,\ell^{'}>\ell).
\end{equation}
Next, by substituting $T$ above into (P1), it can be equivalently derived as
\begin{equation*}\text{(P2)}
\begin{aligned}
& \max\limits_{T,\{b_k\}} \quad T
\\ \text{s.t.} \;\; &\text{(C1)} \;\; b_{k,m}^2 \leq P_k, \quad  \forall(k,m),
\\ &\text{(C2)} \;\; \sum \limits_{m=1}^M b_{k,m}^2 \leq \hat{P}_k, \forall\; k,
\\&\text{(C3)} \;\; T = \sum\limits_{m=1}^M T_{m,\ell,\ell^{'}},\; \forall (\ell,\ell^{'}>\ell),
\\&\text{(C4)} \;\; T_{m,\ell,\ell^{'}} \leq  \dfrac{\left( \hat{\mu}_{\ell,m} - \hat{\mu}_{\ell^{'},m}\right)^2 }{ \hat{\sigma}_{m}^2},\; \forall (\ell,\ell^{'}>\ell,m).
\end{aligned}
\end{equation*}
(P2) is still a non-convex problem. In the sequel, an equivalent d.c. form is first derived and the problem is addressed by the typical method of SCA \cite{razaviyayn2014SCA}. To begin with, by substituting $\hat{\mu}_{\ell,m}$, $\hat{\mu}_{\ell^{'},m}$ and $\hat{\sigma}_{m}^2$ in \eqref{Eq:ReceiveFeatureDistribution} into the fourth constrain of (P2) and with some simple derivations, we can get 
\begin{equation}\label{Eq:EquivalentConstraint}
\begin{aligned}
    & \left(\sum \limits_{k=1}^K h_k b_{k,m} \right)^2  \frac{\left( {\mu}_{\ell,m} - {\mu}_{\ell^{'},m}\right)^2}{T_{m,\ell,\ell^{'}}} 
    \\ \geq & \left(  \sum\limits_{k=1}^K h_k b_{k,m} \right)^2 \sigma_{m}^2 + \sum\limits_{k=1}^K h_k^2 b_{k,m}^2 \delta_{k,m}^2 + \delta_{0}^2,
\end{aligned}
\end{equation}
for all $(\ell,\ell^{'}>\ell,m)$. In \eqref{Eq:EquivalentConstraint}, it is observed that both sides of the inequality are convex. Define the left-side term as 
\begin{equation*}
    Q_{m,l,l^{'}}\left( \{ b_{k,m} \},T_{m,\ell,\ell^{'}} \right) = \left(\sum \limits_{k=1}^K h_k b_{k,m} \right)^2 \frac{\left( {\mu}_{\ell,m} - {\mu}_{\ell^{'},m}\right)^2}{T_{m,\ell,\ell^{'}}},
\end{equation*}
for all $(\ell,\ell^{'}>\ell,m)$. It is no less than its first-order Taylor expansion at $\left( \{ b_{k,m}^{[t]} \},T_{m,\ell,\ell^{'}}^{[t]} \right)$, which is denoted as $\hat{Q}_{m,\ell,\ell^{'}} ^{\left[ t \right]} \left( \{ b_{k,m} \},T_{m,\ell,\ell^{'}} \right)$. That says, $Q_{m,\ell,\ell^{'}}\left( \{ b_{k,m} \},T_{m,\ell,\ell^{'}} \right)
 \geq  \hat{Q}_{m,\ell,\ell^{'}} ^{\left[ t \right]} \left( \{ b_{k,m} \},T_{m,\ell,\ell^{'}} \right)$. 
 
Thereby, the method of SCA can be utilized to solve (P2) via iteratively solving the following approximated convex problem, say (P3), by using  solution in the last iteration as the reference point of the first-order Taylor expansion. 
\begin{equation*}\text{(P3)}\quad
\begin{aligned}
& \max \quad T
\\ \text{s.t.} \;\; &\text{(C1), (C2), (C3)}, 
\\ & \left(  \sum\limits_{k=1}^K h_k b_{k,m} \right)^2 \sigma_{m}^2 + \sum\limits_{k=1}^K h_k^2 b_{k,m}^2 \delta_{k,m}^2 + \delta_{0}^2 
\\ & \leq \hat{Q}_{m,\ell,\ell^{'}} ^{\left[ t \right]} \left( \{ b_{k,m} \},T_{m,\ell,\ell^{'}} \right) ,\quad \forall (\ell,\ell^{'}>\ell,m).
\end{aligned}
\end{equation*}
Besides, (P3) can be solved by the common convex algorithms using e.g., the cvx toolbox. 

\section{Performance Evaluation}
\subsection{Experiment Setup}

\begin{figure}[!t]
	\centering	  
        \includegraphics[width=0.36\textwidth]{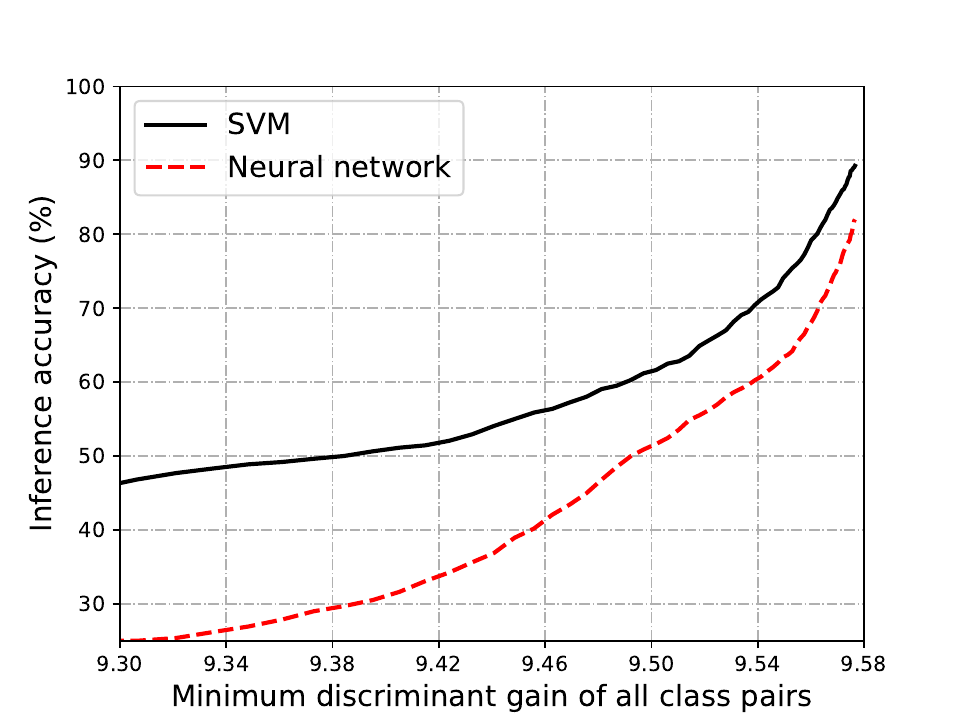}
	\caption{Inference accuracy versus minimum discriminant gain}
	\vspace{-0.4cm}
	\label{fig:disc_acc}
\end{figure}

\begin{figure*}[t]
	\centering
	\subfloat[Inference accuracy with SVM versus number of devices]{\includegraphics[width=0.36\textwidth]{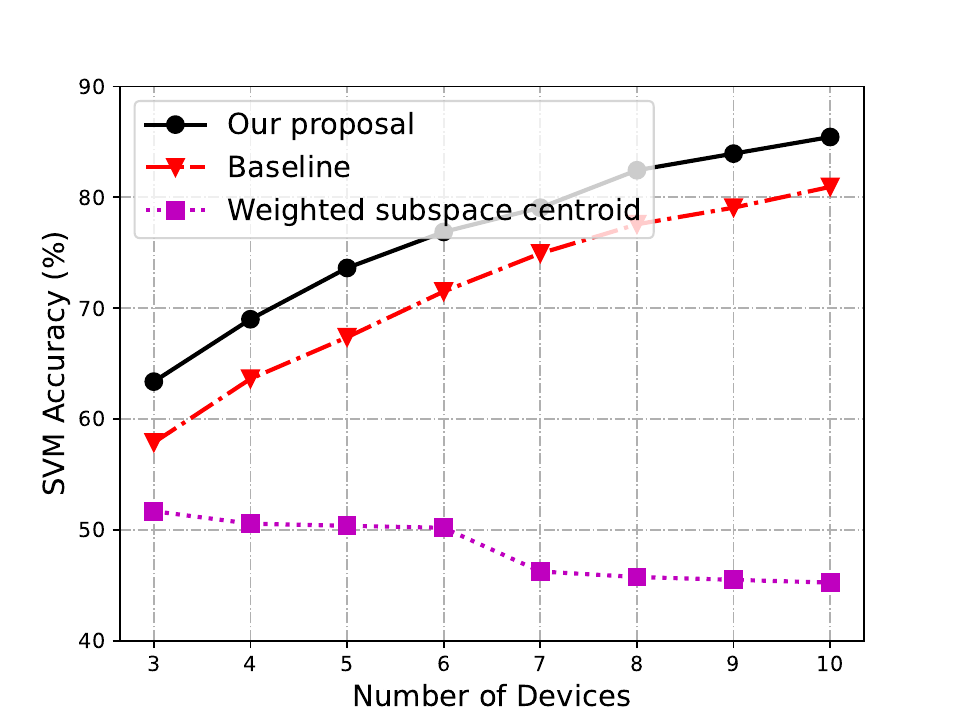}
		\label{fig:svm_device_numbers_accuracy}}\hfil
	\subfloat[Inference accuracy with MLP versus number of devices]{\includegraphics[width=0.36\textwidth]{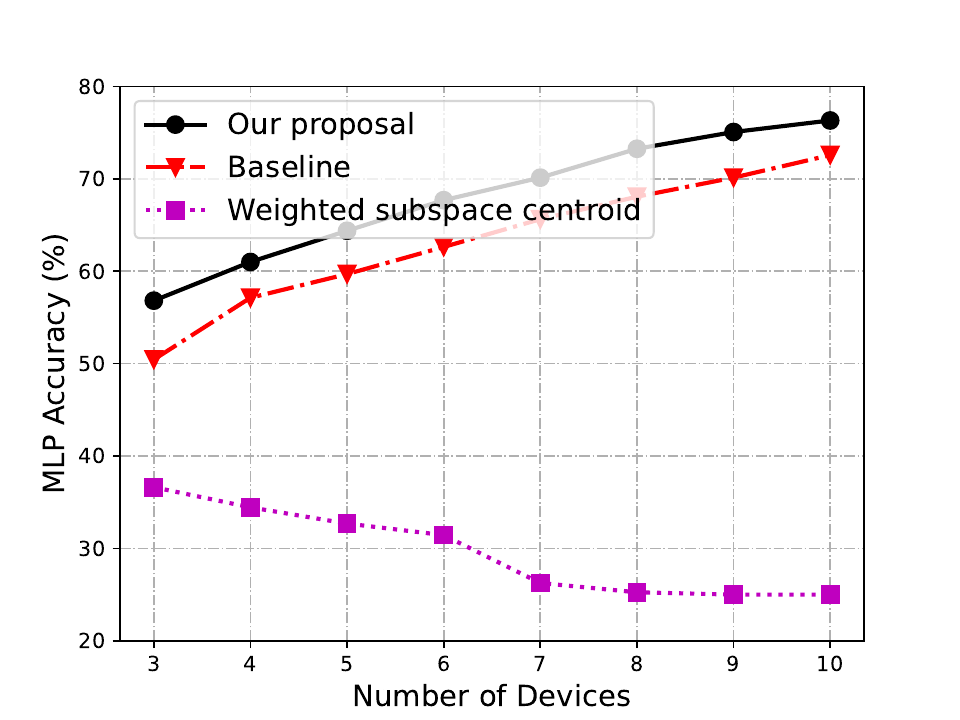}
		\label{fig:mlp_device_numbers_accuracy}}
	\caption{Inference accuracy comparison among different models under different number of devices.}
	\vspace{-0.7cm}
	\label{fig:device_number_accuracy}
\end{figure*}

\begin{figure*}[t]
	\centering
	\subfloat[Inference accuracy with SVM versus transmit power]{\includegraphics[width=0.36\textwidth]{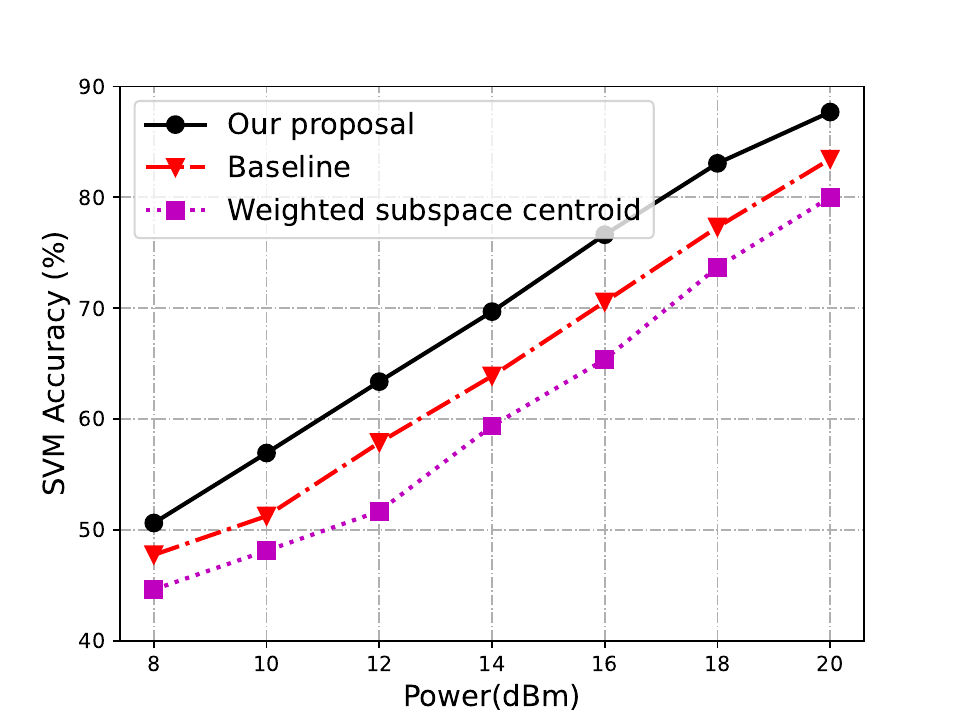}
		\label{fig:svm_power_accuracy}}\hfil
	\subfloat[Inference accuracy with MLP versus transmit power]{\includegraphics[width=0.36\textwidth]{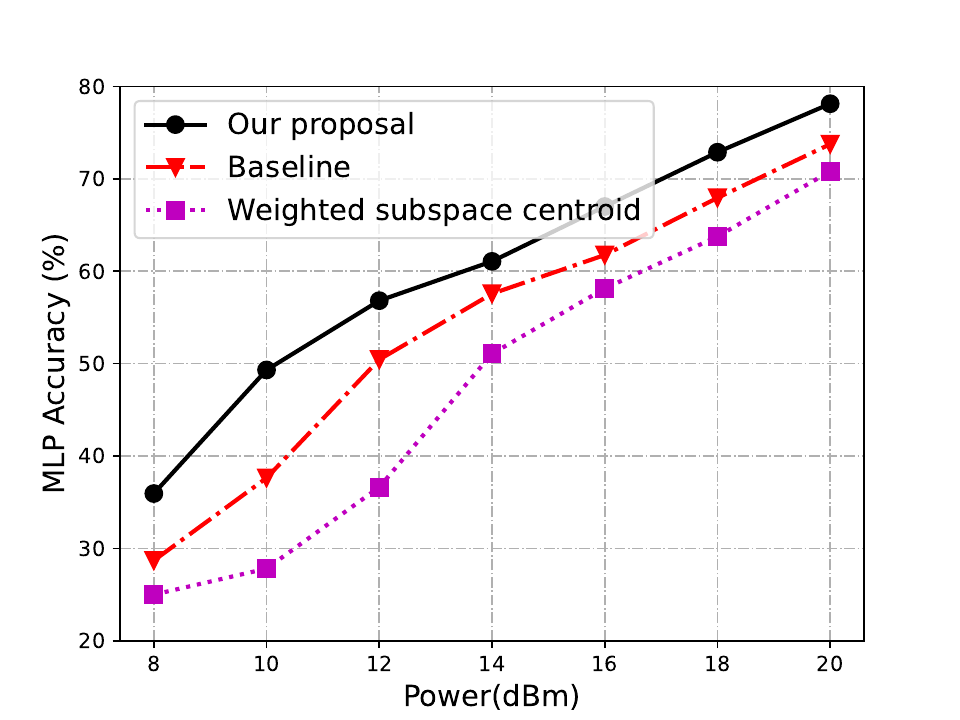}
		\label{fig:mlp_power_accuracy}}
	\caption{Inference accuracy comparison among different models under different transmit power.}
	\vspace{-0.7cm}
	\label{fig:power_accuracy}
\end{figure*}

Consider a single-cell network with a radius of 500 meters, where the server equipped with a single-antenna base station is located at the center and $K$ single-antenna devices are randomly distributed. The channel gain model for each device $k$ is $H_{k} = \left\vert \phi_{k}h_{k} \right\vert^{2}$, where $\phi_{k}$ and $h_{k}$ denote the large-scale and small-scale fading propagation coefficients, respectively. 

A human motion recognition inference task is considered, where there are four classes including child walking, child pacing, adult walking, and adult pacing. A high-fidelity wireless sensing simulator proposed in \cite{Li2021SPAWC} is applied for generating sensory data samples. In the training phase, the model is trained using 6400 data samples to get well-trained AI models at the server. In the inference phase, a noise-corrupted raw sensory data sample is generated for each device, from which local feature vectors are extracted. At the server, all local feature vectors are aggregated for completing the downstream inference task. For testing the inference accuracy, 1600 independent inference experiments are conducted.  Two AI models are adopted for the task. One is a support vector machine (SVM). The other is a multi-layer perceptron (MLP) neural network with two hidden layers, each with 80 and 40 neurons respectively. By default, the number of devices, the feature noise variance, the number of extracted feature dimensions, the transmit power, and the stepsize of SCA are set
as $K=3$, $\delta_{k,m}^2=0.4$, $M=12$, $P_{k}=12 dBm$, and $\alpha=0.7$ respectively, unless specified otherwise. 

For comparison, we consider three schemes as follows.
\begin{itemize}
	\item \textit{Baseline}: All the parameters are allocated following the task-oriented AirComp scheme in \cite{wen2022task}, which aims at maximizing average pair-wise discriminant gain.
	\item \textit{Weighted subspace centroid}: All the parameters are allocated following the traditional AirComp scheme in \cite{Zhu2019IoTJ}, whose design criterion is MMSE.
	\item \textit{Joint optimization of all feature elements (our proposal)}: All parameters are set follow the proposed scheme.
\end{itemize}

\vspace{-0.5em}
\subsection{Experimental Results}

In this part, the experimental results are shown. The relation between inference accuracy and the minimum discriminant gain is first presented, followed by the comparison among the three schemes.

\subsubsection{Inference accuracy v.s. discriminant gain}
Fig. \ref{fig:disc_acc} illustrates the relation between the minimum pair-wise discriminant gain and inference accuracy. 
It shows that the inference accuracy rises as the minimum pair-wise discriminant gain grows for both models. That's because a larger minimum pair-wise discriminant gain leads to a larger distance for arbitrary class pairs in the feature space, hence resulting in larger inference accuracy. 

\subsubsection{Inference accuracy v.s. number of devices}
Fig. \ref{fig:device_number_accuracy} shows the inference accuracy of the SVM and MLP models in terms of various device numbers. From the figure, the performance of the MMSE-based scheme, say the scheme of weighted subspace centroid, decreases with an increasing number of devices, since larger devices lead to a lower MMSE due to the requirement of channel equalization for all devices, as well as MMSE cannot characterize the inference accuracy. The inference accuracy of both task-oriented AirComp schemes grows as the number of devices increases. That's because the diversity of data can be fully utilized and the sensing noise of different devices is adaptively suppressed by allowing different receive power of different devices. More importantly, our proposed scheme has the best performance due to the novel metric of maximizing the minimum discriminant gain of all class pairs and the joint optimization of all feature elements.

\subsubsection{Inference accuracy v.s. transmit power}
Fig. \ref{fig:power_accuracy} presents the inference accuracy of the SVM and MLP models under various transmit power levels. For both models, the inference accuracy of all schemes increases as transmit power rises, since more transmit power can suppress the channel noise without doubt. For similar reasons, our proposed scheme has the best performance.

The experimental findings show that the proposed scheme has the best performance and verify our theoretical analysis.

\section{Conclusion}
This work proposed a task-oriented AirComp scheme for edge-device co-inference, which addresses the issue of unbalanced classification accuracy by maximizing the minimum pair-wise discriminant gain and jointly optimizing the long-term transmission over multiple time slots. It opens several directions like extending this scheme to the multiple-input multiple-output systems, using unlicensed spectrum \cite{7557056}, the scenario of the large language model, etc.
Particularly, since AirComp has been shown a promising technique for implementing federated learning (see e.g., \cite{du2023JSAC-FEELscheduling}), task-oriented AirComp for federated learning is an attractive research direction.


\bibliographystyle{IEEEtran}
\bibliography{reference}

\end{document}